%% file: main.tex
\documentclass[
 reprint,
 amsmath,amssymb,
 aps,
 prl,
 superscriptaddress,
]{revtex4-1}

\usepackage{graphicx}
\usepackage{dcolumn}
\usepackage{bm}
\usepackage[colorlinks]{hyperref}

\begin{document}

\preprint{APS/123-QED}

\title{Measurements of Nonequilibrium Interatomic Forces in Photoexcited Bismuth}

\author{Samuel W. Teitelbaum}

\affiliation{PULSE Institute of Ultrafast Energy Science, SLAC National Accelerator
Laboratory, Menlo Park, California 94025, USA}

\affiliation{Stanford Institute for Materials and Energy Sciences, SLAC National
Accelerator Laboratory, Menlo Park, California 94025, USA}
\email{steitelb@slac.stanford.edu}


\author{Thomas C. Henighan}

\affiliation{PULSE Institute of Ultrafast Energy Science, SLAC National Accelerator
Laboratory, Menlo Park, California 94025, USA}

\affiliation{Department of Physics, Stanford University, Stanford, California
94305, USA}

\author{Hanzhe Liu}

\affiliation{PULSE Institute of Ultrafast Energy Science, SLAC National Accelerator
Laboratory, Menlo Park, California 94025, USA}

\affiliation{Department of Physics, Stanford University, Stanford, California
94305, USA}

\author{Mason P. Jiang}

\affiliation{Department of Physics, Stanford University, Stanford, California
94305, USA}

\affiliation{PULSE Institute of Ultrafast Energy Science, SLAC National Accelerator
Laboratory, Menlo Park, California 94025, USA}

\author{Diling Zhu}

\affiliation{Linac Coherent Light Source, SLAC National Accelerator Laboratory, Menlo Park, California
94025, USA}

\author{Matthieu Chollet}

\affiliation{Linac Coherent Light Source, SLAC National Accelerator Laboratory, Menlo Park, California
94025, USA}

\author{Takahiro Sato}

\affiliation{Linac Coherent Light Source, SLAC National Accelerator Laboratory, Menlo Park, California
94025, USA}

\author{\'Eamonn D. Murray}
\affiliation{Department of Physics and Department of Materials, Imperial College
London, London SW7 2AZ, United Kingdom}

\author{Stephen Fahy}
\affiliation{Tyndall National Institute, Cork, Ireland}
\affiliation{Department of Physics, University College Cork, Cork, Ireland}

\author{Shane O'Mahony}
\affiliation{Tyndall National Institute, Cork, Ireland}
\affiliation{Department of Physics, University College Cork, Cork, Ireland}

\author{Trevor P. Bailey}
\affiliation{Department of Physics, University of Michigan, Ann Arbor, Michigan
48109, USA}

\author{Ctirad Uher}
\affiliation{Department of Physics, University of Michigan, Ann Arbor, Michigan
48109, USA}

\author{Mariano Trigo}
\affiliation{PULSE Institute of Ultrafast Energy Science, SLAC National Accelerator
Laboratory, Menlo Park, California 94025, USA}
\affiliation{Stanford Institute for Materials and Energy Sciences, SLAC National
Accelerator Laboratory, Menlo Park, California 94025, USA}

\author{David A. Reis}
\affiliation{PULSE Institute of Ultrafast Energy Science, SLAC National Accelerator
Laboratory, Menlo Park, California 94025, USA}
\affiliation{Stanford Institute for Materials and Energy Sciences, SLAC National
Accelerator Laboratory, Menlo Park, California 94025, USA}
\affiliation{Department of Applied Physics, Stanford University, Stanford, California
94305, USA}
\affiliation{Department of Photon Science, Stanford University, Stanford, California
94305, USA}
\begin{abstract}
We determine experimentally the excited-state interatomic forces in photoexcited bismuth. The forces are obtained by a constrained least-squares fit of the excited-state dispersion obtained by femtosecond time-resolved x-ray diffuse scattering to a fifteen-nearest neighbor Born-von Karman model. We find that  the observed softening of the zone-center $A_{1g}$ optical mode and transverse acoustic modes with photoexcitation are primarily due to a weakening of three nearest neighbor forces along the bonding direction.  This provides a more complete picture of what drives the partial reversal of the Peierls distortion previously observed in photoexcited bismuth.
\end{abstract}

\date{\today}

\maketitle


In light-driven nonequilibrium systems, the interatomic forces can be strongly modified, either directly by creation of hot carriers  \citep{Fritz2007,Beaud2014,Schmitt2008,Lindenberg2005}, or with  low-energy resonant excitations \citep{Forst2011,Basov2017} revealing hidden states and exotic phenomena.  However, characterizing interatomic forces in these transients is challenging.  Typically, only zone-center (zero-wavevector) modes are probed in the nonequilibrium state, meaning most of the degrees of freedom are experimentally inaccessible, and thus information about the microscopic nonequilibrium interatomic forces is limited.   In equilbrium, these forces can be obtained by fitting the phonon dispersion relation \citep{Born1954} derived from a combination of experimental probes, such as inelastic neutron scattering (INS), inelastic x-ray scattering (IXS), Raman scattering, and infrared scattering, to  particular ground-state force models \citep{Brockhouse1955,Cowley1964,Baron2014,Baron2014a,Krisch2007}. 

Time- and momentum-resolved x-ray diffuse scattering \citep{Trigo2013} probes the \emph{nonequilibrium} phonon dispersion relation throughout the Brillouin zone.  To date, time-resolved diffuse scattering has measured near-equilibrium dispersion relations \citep{Zhu2015}, mode mixing \citep{Jiang2016}, and three-body decay \citep{Teitelbaum2018}.  In this work, we demonstrate that x-ray scattering can be used to extract transient excited-state interatomic forces in the prototypical Peierls-distorted material bismuth, using a combination of time-resolved Bragg scattering and time-resolved diffuse scattering at varying laser excitation levels.  We show that changes to the phonon dispersion are driven primarily by softening of the nearest-neighbor bonds, consistent with photoexcitation partially reversing the Peierls distortion.

At ambient temperature and pressure, bismuth has a Peierls-distorted structure with a primitive basis of two atoms per unit cell.  This structure can be thought of as a rhombohedral distorted  simple cubic structure, with three nearest-neighbor bonds connecting atoms into  layers of buckled chains orthogonal to the trigonal (i.e. $(1\ 1\ 1)$ in the rhombohedral basis) direction. Photoexcitation weakens the interatomic forces and drives atoms towards a more symmetric sixfold-coordinated structure, along the $A_{1g}$ mode coordinate \citep{Murray2007a,Murray2005a,Johnson2009,Cheng1990}.  This motion can be observed in time-resolved optical reflectivity \citep{Cheng1990,Teitelbaum2018b,Hase1998}, and in x-ray diffraction \citep{Fritz2007,Sokolowski-Tinten2003}.  However, a more complete picture of the renormalization of nonequilibrium atomic forces requires measurements of the excited-state phonon dispersion \citep{Murray2007a}.  


\begin{figure*}
    \centering
    \includegraphics{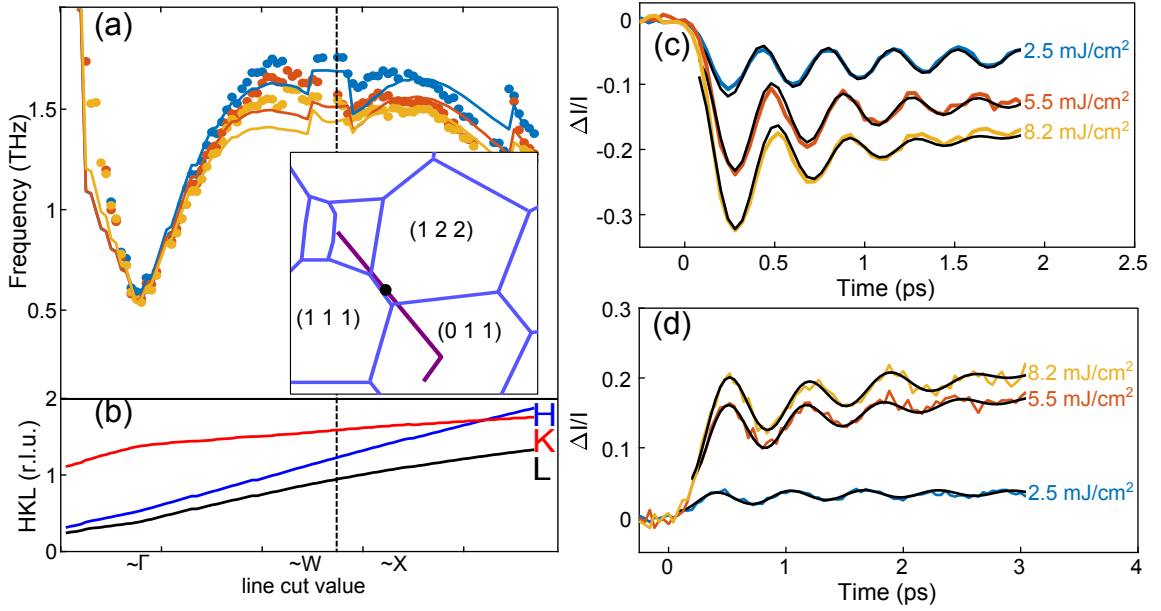}
    \caption{(color online) (a) Measured frequencies of phonon mean-square displacements for photoexcited bismuth as a function of incident laser fluence.  The data correspond to the wavevectors in panel (b).  The inset in panel (a) shows a portion of reciprocal space subtended by the detector. The  corresponding path (purple) and the Brillouin zone boundaries (blue) are also indicated.  (c) Time-dependent relative change in the Bragg  intensity of the (2 2 3) peak, showing oscillations of the $A_{1g}$ mode about its new quasi-equilibrium position.  (d) Time-dependent relative change in the diffuse scattering intensity at the wavevector Q=(1.00 1.61 1.39) (black dots in the inset to (a), dashed line in (a) and (b).  The solid lines through the data in (a), (b) and (c) correspond to the best fits as described in the text.}
    \label{fig:Data_Overview}
\end{figure*}

To extract the excited-state dispersion relation, we performed time-resolved x-ray scattering experiments at the X-ray pump-probe (XPP) instrument at the Linac Coherent Light Source (LCLS) x-ray free electron laser with a photon energy of 9.5 keV  \citep{Chollet2015}. The experimental details were similar to that used in ref. \citep{Teitelbaum2018} to measure phonon-phonon coupling. 
The sample was a 50 nm thick epitaxial film of bismuth with its surface normal along the  (111) direction.  The crystal was rotated such that the x rays propagated at a 0.5 deg. grazing angle from the surface and at a 71 deg. angle with respect to $(2 \bar{1} \bar{1})$ (binary axis). The incident fluence of the 800 nm p-polarized, 65 fs  excitation laser was varied between $2.5 - 8.0 \ \mathrm{mJ/cm^2}$. A pixel array detector (CSPAD \citep{Herrmann2013}) was used to simultaneously collect scattered x rays over a wide range of momentum transfer $\bm{Q}$. 



The procedure for obtaining the phonon frequencies from the time-resolved scattering data is as follows:  first, the detector data was interpolated to a 256 by 256 spatial and 152 temporal array. Each spatial point on the array was fit to a sum of exponential decay and decaying cosine functions by linear prediction fitting \cite{BARKHUIJSEN1985} with three damped harmonic oscillators \footnote{See supplemental information for a map of the fitted frequencies on the detector, the fit convergence as a function of the number of fit parameters.}.  The fitted frequency shown in fig. \ref{fig:Data_Overview}(a) is the oscillatory component between 0.1 and 2.5 THz with the largest amplitude.  Away from the Bragg peaks, the observed frequency corresponds to twice the phonon frequency, as expected for oscillations in the variance of the phonon displacements \citep{Trigo2013,Henighan2016}.  To leading order, the diffuse scattering intensity in high-quality crystals is proportional to the variance of the phonon displacements. Thus, the oscillations in the diffuse scattering intensity correspond to two-phonon excitations driven by the change in interatomic forces upon photoexcitation.  

The data are fit to a phonon dispersion derived from a model for the interatomic force constants (IFCs) as follows. For pair-wise potentials in the harmonic approximation,  the interatomic forces  $\Phi$ are the second derivatives of the total energy $E$ with respect to the atomic displacements $u_{i,\alpha}$

\begin{equation}
    \Phi_{ij}^{\alpha,\beta} = \frac{\partial^2 E}{\partial u_{i,\alpha}\partial u_{j,\beta}} \label{eq:IFC}
\end{equation}

Each $\Phi_{ij}^{\alpha,\beta}$ corresponds to the force on atom $j$ in direction $\beta = (x,y,z) $ produced by a displacement of atom $i$ in direction $\alpha$.  The phonon dispersion is obtained from the dynamical matrix \citep{Born1954}, which is the Fourier transform over the force constants

\begin{equation}
    D_{ij}^{\alpha,\beta}(\bm{q}) = \frac{1}{{M}}\sum_{ij} \Phi_{ij}^{\alpha,\beta} e^{-i \bm{q} \cdot (\bm{r_i-r_j})}. \label{eq:dyn_matrix}
\end{equation}

The factor $M$ is the mass of the bismuth atom.  The eigenvalues and eigenvectors of the dynamical matrix at wavevector $\bm{q}$ are the square of the harmonic phonon normal mode frequencies $\omega^2_i(\bm{q})$, and phonon polarization $\bm{\epsilon_{i}}(\bm{q})$, respectively. 

We limit the force constants to a $4\times 4\times 4$ supercell, corresponding to 1152 force constants for 128 atom pairs.  Additional symmetries of the crystal reduce this number to 21 independent $3\times3$ force matrices.  Furthermore, we assume that the photoexcitation does not change the overall bonding directions of the system significantly.  The assumption implies the eigenvalues, but not the eigenvectors, of each force matrix change, further reducing the adjustable forces to three per symmetry-inequivalent atom pair.  This fitting procedure is equivalent to a reduced Born-von Karman model \citep{Born1954} where the eigenvectors of the interatomic forces are known.

\begin{figure}
    \centering
    \includegraphics{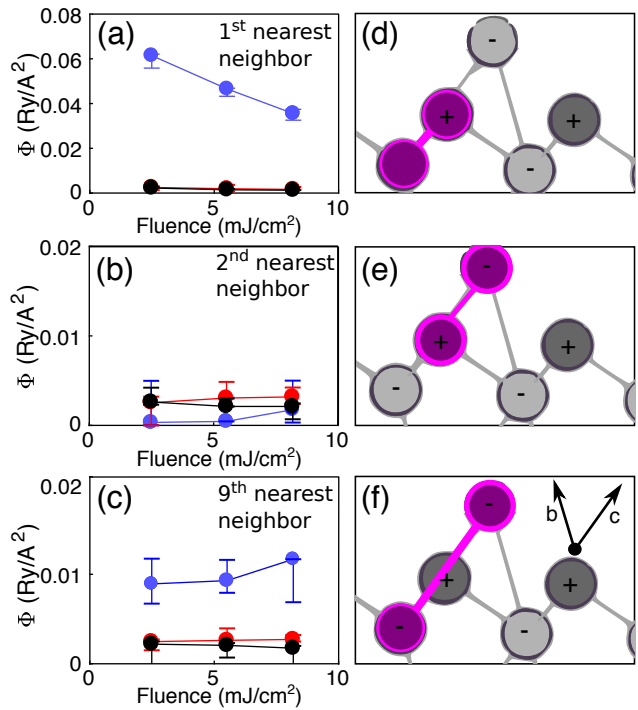}
    \caption{(color online) (a)-(c) Eigenvalues for the force matrices of the three largest interatomic forces, sorted by interatomic distance. (d)-(f) Illustration of the eigenvectors and atom pairs for the forces in (a)-(c) in the (b,c) plane.  Dark and light gray atoms are the first and second atoms in the basis, respectively.  The $(+)$ and $(-)$ symbols indicate the out of plane positions.  (f) shows the primitive lattice vectors in the plane.}
    \label{fig:IFC_results}
\end{figure}

The eigenvectors of the real-space force matrix correspond to the directions where the restoring force applied is parallel to the atom displacement.  Typically, this corresponds to one component along the bond axis, one component in the bonding plane, and one component out of plane.  Of the 21 unique force matrices, we will discuss the three with the largest trace, which correspond to the nearest-neighbor, second-nearest neighbor, and ninth-nearest neighbor forces.  The other forces are kept at the values obtained by density functional perturbation theory (DFPT) calculations.  The direction and network of the bonds is shown in fig. \ref{fig:IFC_results}(d)-(f).

We incorporate one more piece of experimental information into our model: the frequency of the zone-center $A_{1g}$ mode observed near the Bragg peaks.  This is analogous to including Raman scattering results as a constraint in fitting INS or IXS data to a dispersion \citep{Cowley1964}.  The $A_{1g}$ frequency was incorporated as an additional term in the optimization function, and was weighted as highly as the total mean-square error from the acoustic phonon branches.  

For each incident laser fluence, the initial guess was the full force constants derived from DPFT calculations.  Each detector pixel is sensitive to six possible phonon branches, but the signal is strongest for the lower frequency modes, and thus only one of the acoustic phonon branches is typically observed.  The second harmonic of the LO and TO phonon branches were not observed in our experiment. The observed oscillation frequency at each pixel was assigned to a particular phonon branch by the highest computed thermal diffuse scattering (TDS) intensity \citep{Ruqing2005,Walker1956}.  This assignment assumes similar excitation amplitude relative to equilibrium for all phonon modes, such that the oscillation amplitude is proportional to the initial mean-square displacement of the mode.  The TDS intensity for each branch was computed in a similar manner to ref. \citep{Ruqing2005}. For most pixels used in this fit, the fitted branch corresponds to one of the transverse acoustic modes.  

For the purposes of the fitting, we found that the use of the three interatomic force matrices described above was sufficient to minimize the reduced $\chi^2$ to approximately 300.  Introducing additional adjustable forces to the fitting did not significantly improve the fit.  Refer to the supplemental information for additional details about the fitting procedure, optimization function, model, and fit convergence.

A reduced $\chi^2$ of 300 implies that the uncertainty in our fit parameters is not dominated by statistics in the fitted frequency values, but rather by systematic uncertainties and/or the ability of our model to fully capture the observed phonon dispersion.  Systematic uncertainties in this experiment arise predominantly from the geometry calibration, which introduces an uncertainty in the mapping of each detector pixel to reduced a wavevector $\bm{q}$.  Therefore, systematic uncertainties in the forces are estimated by varying the crystal alignment $\pm 1$ degree along the (1 1 1) direction in the fitting routine and performing the fit multiple times.  The error bars shown in fig. \ref{fig:IFC_results}(a)-(c) are the largest and smallest fit results produced by the fitting routine over the range of possible geometries.


\begin{figure}
    \centering
    \includegraphics{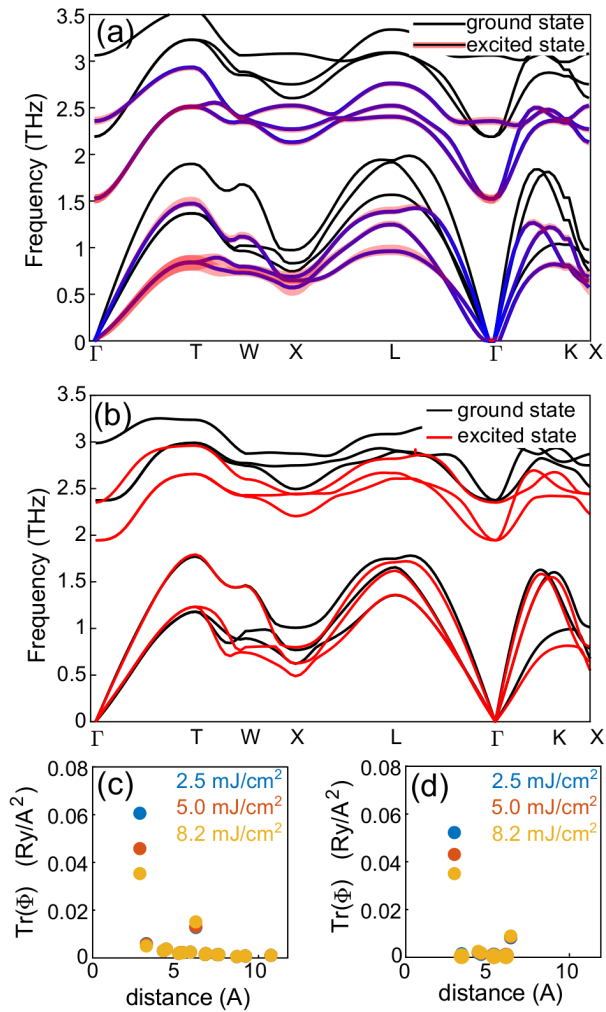}
    \caption{ (color online) (a) Reconstructed dispersion along the high-symmetry directions in bismuth.  Black lines show the (extrapolated) ground-state dispersion, blue lines show the photo-excited dispersion at 2.5 mJ/cm$^2$.  Red shaded areas indicate systematic uncertainties in the dispersion relation as described in the text.  (b) DPFT calculations of the ground-state and excited-state forces for the same excitation level as in (a). (c) and (d) show the trace of the interatomic force matrices as functions of distance and fluence from the reconstructed forces and DPFT calculations, respectively.   The forces at 2.5 and 7 A interatomic distance are the bonds shown in fig. \ref{fig:IFC_results}(d) and (f), respectively.}  
    \label{fig:forces-followup-figure}
\end{figure}

The best fit to the 1st, 2nd, and 9th nearest-neighbor excited-state forces are shown in fig. \ref{fig:IFC_results} as a function of fluence.  Panels (a)-(c) show the eigenvalues of the three force matrices.  The first-nearest-neighbor force, shown in panels (a) and (d), has the largest change with fluence.  This is dominated by the first eigenvalue, which corresponds to the restoring force along the bonding direction.  Intuitively, these bonds can be thought of as those responsible for the Peierls distortion (dimerization of Bi atoms, accompanied by opening of a gap at the Fermi surface).  When this bond is sufficiently weakened, the crystal transitions into a high-symmetry phase \citep{Teitelbaum2018a,Murray2005a}.  At 8 mJ/cm$^2$, this force is already weakened to almost 50\% of its ground-state value.  As a result, at this fluence, the $A_{1g}$ mode is softened from 2.95 to 2 THz (a 30\% reduction in freuency).  This force constant dominates the frequency of the $A_{1g}$ mode.

In the force matrix with the second largest trace (fig. \ref{fig:IFC_results}(c) and (f)), the forces along the bonding direction stiffen slightly with increasing fluence, though it is within our experimental uncertainty.  The force matrix with the third largest trace, while improving the fit, does not change appreciably with increasing fluence within the uncertainties of our fitting procedure.  Modifying additional forces beyond the third and fourth strongest do not significantly improve the fit quality.  

The reconstructed dispersion along high-symmetry directions is shown in fig. \ref{fig:forces-followup-figure}(a).  The shaded areas indicate uncertainties in the excited-state dispersion based on systematic propagation of the uncertainties in the extracted forces.  We note that the softening of the acoustic modes is substantially more pronounced than that predicted by the constrained density functional theory calculations in ref. \cite{Murray2007a}, especially for the TA mode near the L point.  The values of the trace in the forces as a function of distance and incident fluence is shown in fig. \ref{fig:forces-followup-figure}(b).  This shows the largest force, and largest force change with increasing laser fluence, is the nearest-neighbor force shown in fig. \ref{fig:IFC_results}(a) and (d).  

This finding is in disagreement with predictions in ref. \citep{Murray2007a}, where DFT calculations demonstrated a small increase in the ninth-nearest neighbor force.  Results from our DPFT calculations are shown in fig. \ref{fig:forces-followup-figure}(b).  The excited-state forces are extrapolated such that the $A_{1g}$ frequency matches that of the experimental data.  It can be seen that other than along the $\Gamma$-X direction, the pronounced softening seen in the reconstructed dispersion is not fully reproduced by DPFT calculations.  This disagreement could explain the more pronounced softening we observe in the reconstructed dispersion compared to DFT calculations.

Note that the largest eigenvalue of the nearest-neighbor force is nearly five times larger than all other fit parameters, and dominates both the $A_{1g}$ mode softening and the softening of the observed TA branches.  This force is the binding force of the dimer chains, and is the force largely responsible for the Peierls distortion.  Intuitively, when the force becomes equal to the 3rd largest force (next-nearest-neighbor), the Peierls distortion vanishes, as there is an equal force from either dimerization direction.  Extrapolating the force constant to high fluence, we note that a linear extrapolation would result in the force constant going to zero at approximately 16 mJ/cm$^2$ incident laser fluence.  This is approximately the expected fluence for when an electronically driven Peierls distortion in bismuth vanishes, and the excited-state lattice structure is symmetric \citep{Teitelbaum2018b,Murray2005a}.

In conclusion, we have demonstrated a method for determining the excited-state interatomic forces  using wavevector-resolved femtosecond x-ray scattering. These time-domain experiments enable high-resolution measurements of the phonon frequencies in the transient state.  In the case of  bismuth, we determine for the first time, the connection between photoexcitation and the interatomic forces that drive the Peierls distortion. This approach is applicable to a diverse range of light-driven materials, e.g. \citep{Basov2017,Forst2011,Mitrano2016}.

\begin{acknowledgments}
This work was supported by the U.S. Department of Energy, Office of Science, Office of Basic Energy Sciences through the Division of Materials Sciences and Engineering under Contract No. DE-AC02-76SF00515. C. Uher and T. Bailey acknowledge support from the Department of Energy, Office of Basic Energy Science under Award \# DE-SC-0008574. Work at the Tyndall National Institute was supported by Science Foundation Ireland award 12/IA/1601 and Irish Research Council award GOIPG/2015/2784. Measurements were carried out at the Linac Coherent Light Source, a national user facility operated by Stanford University on behalf of the U.S. Department of Energy, Office of Basic Energy Sciences. Preliminary measurements were performed at the Stanford Synchrotron Radiation Lightsource (Beamline 7\textendash 2), SLAC National Accelerator Laboratory.
\end{acknowledgments}

\bibliographystyle{apsrev4-1}
\bibliography{Bi_TA_softening_references.bib}

\clearpage
\onecolumngrid

\input{supplement.tex}

\end{document}

%% file: supplement.tex



\title{Supplemental Information For: Measurements of Nonequilibrium Interatomic Forces in Photoexcited Bismuth}

\author{Samuel W. Teitelbaum}

\affiliation{PULSE Institute of Ultrafast Energy Science, SLAC National Accelerator
Laboratory, Menlo Park, California 94025, USA}

\affiliation{Stanford Institute for Materials and Energy Sciences, SLAC National
Accelerator Laboratory, Menlo Park, California 94025, USA}
\section{Supplemental Information}

The supplemental information provides additional information about the fitting routine, the extraction of the interatomic forces from the geometry, and the time and frequency-domain datasets.

\subsection{Fitting Routine for Phonon Frequencies}

For each pixel, linear prediction fitting (LPF) using six components returns one to two values with nonzero frequency.  Linear prediction fitting fits a time-domain signal $S(t)$ to a sum of M exponentially damped oscillators

\begin{equation}
    S(t) = \sum_{i=1}^M A_i e^{-\Gamma_i t}\cos\big(2\pi f_i t + \phi_i \big)
\end{equation}

The number of oscillators used in the LPF fitting is half the number of components with nonzero frequencies (because each component with nonzero frequency contains a sine and cosine component).  Therefore, LPF with six components includes two exponential decays, and up to two oscillating components (with sine and cosine amplitudes).  Of the two oscillating components, we assign any components between 0.1 and 2 THz to an acoustic mode.  This eliminates the contributions of the $A_{1g}$ mode at 2.5-3 THz (depending on the excitation fluence) and the zero and low-frequency components associated with a quasi-DC increase in scattering intensity.

Statistical uncertainties in the fitted frequency as a function of reduced pixel were obtained by performing the fitting routine above for three subsets of the lowest fluence data, with each subset containing 30 experimental runs each out of the 120 total runs used in the final analysis.  The uncertainty is defined as half-width between the largest and smallest fitted frequencies.

\begin{figure*}
    \centering
    \includegraphics{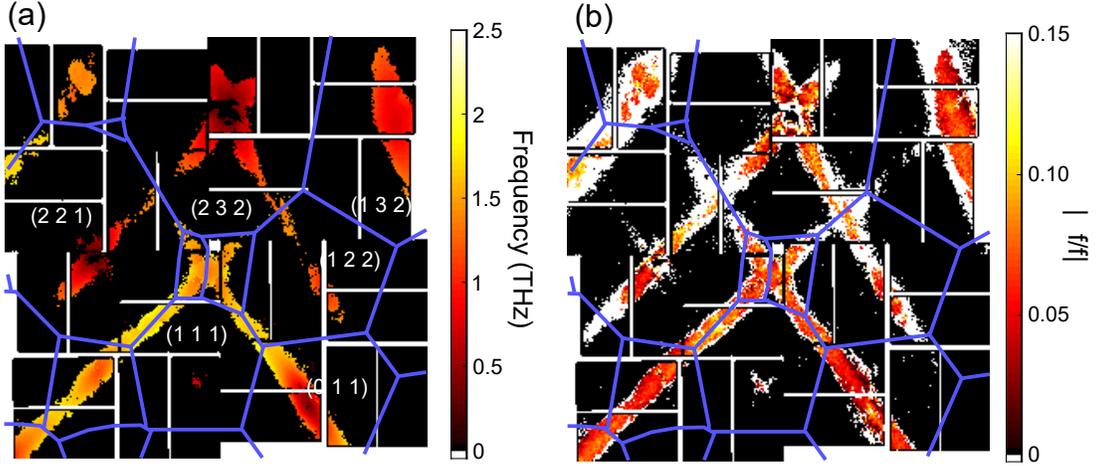}
    \caption{(a) Dominant frequencies extracted from linear prediction fitting for every pixel where the maximum amplitude of the FFT frame passed the threshold of 0.1 (A.U.).  Pixels in black are not fit.  (b) Relative frequency change $\Delta f/f$ as a function of pixel.}
    \label{fig:Bi_area_fitting_results}
\end{figure*}

The extracted frequencies as over the detector image are shown in fig. \ref{fig:Bi_area_fitting_results}(a).  Selected Brillouin zone indices are also labeled.  

\subsection{Fitting Routine for Interatomic Forces}

\begin{figure*}
    \centering
    \includegraphics{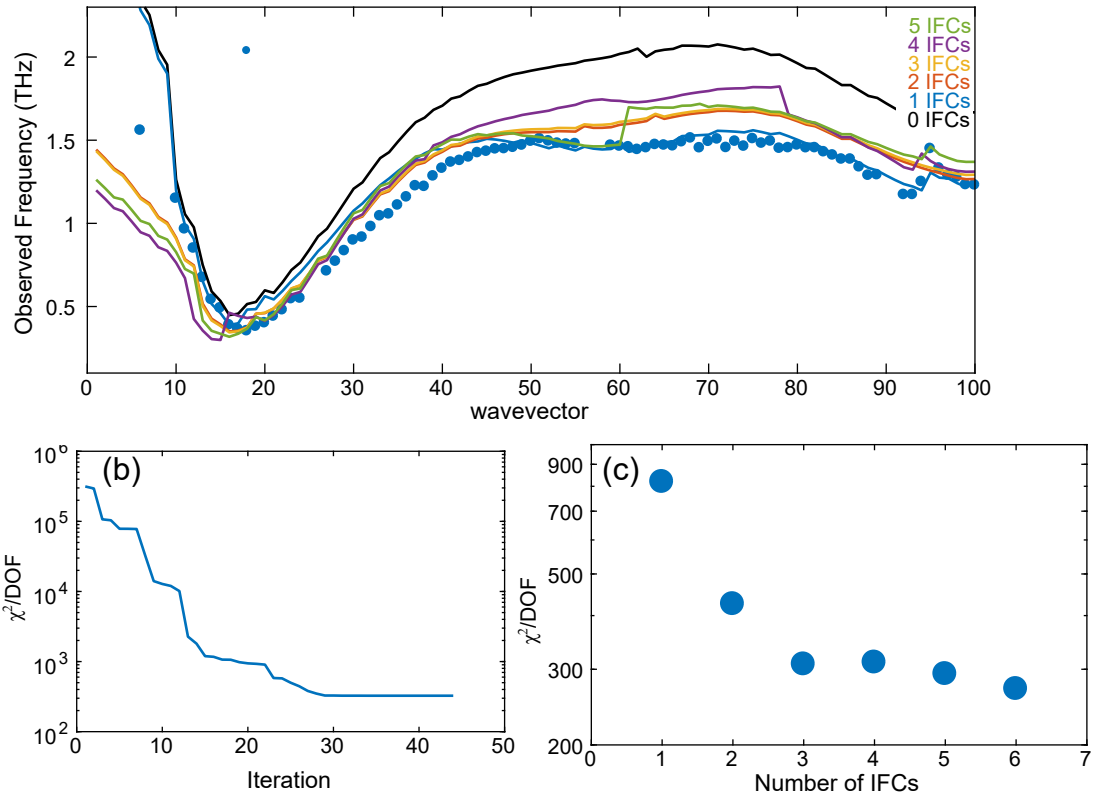}
    \caption{(a) Fitting of the lineout shown in fig. 1 of the main text as a function of the number of IFC forces used.  The black line is the results of the ground-state  DFT calculation with no adjustable parameters.  (b) Reduced $\chi^2$ as a function of the iteration to minimize the mean-square error.  The fitting converges after approximately 50 iterations.  (c) Reduced $\chi^2$ as a function of the number of adjustable IFCs (log y scale).  Adjusting IFCs beyond the fourth does not significantly improve the fit.  For reference, fitting the dispersion with no adjustable parameters is equivalent to the beginning of the fitting iteration in (b), with a reduced $\chi^2$ of $3\times10^5$.}
    \label{fig:Fitting_Convergence}
\end{figure*}

The model is fit by optimizing the adjustable force eigenvalues, which are defined as the vector $X$, with $3N_{IFC}$ components, where $N_{IFC}$ is the number of force matrices used.  The fit result is the vector $X$ that minimizes the least-squares error (eq. \ref{eq:fitting-MSE}) for all detector pixels for which the linear prediction fitting converges.  The least-squares fitting of the observed frequency $f_n$ for all $N$ pixels $n$ on the detector with an observed frequency  is the optimized reduced mean-square error (MSE),

\begin{equation}
    \chi^2(\bm{X}) = \frac{1}{N-n_X}\sum_{n=1}^N \Big(\frac{ 2f^{(fit)}_{n}(\bm{X}) - f^{(obs)}_{n}}{\sigma_{fn}  }\Big)^2 + \Big(\frac{f_{A1g}^{(fit)}(\bm{X})- f^{(obs)}_{A1g}}{\sigma_{fA1g}}\Big)^2. \label{eq:fitting-MSE}
\end{equation}

The first term in eq. \ref{eq:fitting-MSE} is the weighted mean-square error of the second harmonic of the computed phonon frequencies $2f^{(fit)}_{n}$, compared to the experimentally observed frequencies $f^{(obs)}_{n}$.  The factor of two is present because the squeezing oscillations are observed at the second harmonic of the phonon frequency.  The second term in the equation is the mean-square error of the $A_{1g}$ mode frequency. We then minimize the mean-square error $\chi^2$ as a function of the interatomic force parameters $\bm{X}$.    

When fitting the interatomic forces to a model, we must select the number of forces to vary.  Our approach is to select the minimum number of forces to produce a reasonable convergence, in order to minimize any over-fitting or correlation between fit parameters.  Therefore, we fit the cut of the dispersion shown in fig. \ref{fig:Bi_area_fitting_results}(a) to three sets of forces (nine fit parameters).  The fitting results along the line cut shown in fig. \ref{fig:Data_Overview}(a) in the main text as a function of the number of adjustable forces used in the fit are shown in fig. \ref{fig:Fitting_Convergence}(a).  The  dispersion produced from DFT calculations is shown in black (0 changed IFCs). The phonon dispersion produced by DFT calculations is higher frequency than the reconstructed dispersion across the Brillouin zone.  By adjusting the nearest-neighbor forces,  better agreement between the calculated and measured TA branch frequencies is reached.  We can quantify this agreement by observing the mean-square error as a function of the number of interatomic force constant (IFC) matrices modified in the fitting, shown in fig. \ref{fig:Fitting_Convergence}(c).  Because the number of points fitted is much greater than the number of parameters used in the fit, information-theoretic approaches to determining the number of parameters are not appropriate for selecting the number of forces to fit.

The start point for the fitting is the ground-state dispersion relation for all forces. The minimization problem is solved by the constrained nonlinear optimizer algorithm in MATLAB.  The algorithm converges to a minimum value in approximately 50 iterations.  The descent of the algorithm to the minimized value is shown in fig. \ref{fig:Fitting_Convergence}(b) for three IFCs.  

The phonon dispersion is computed for each wavevector $Q$ where the fitting converges, which is 7,080 pixels of the 65,536-pixel reduced pixel array. For each iteration, the dispersion is computed with the adjusted IFCs.  The branches are then assigned by computing the diffuse scattering intensity of all six phonon branches, and taking the maximum for each pixel.  The weighted mean-squared error is then assigned, and added to the mean-squared error from the $A_{1g}$ mode.  A general description for how to compute the diffuse scattering intensity for each phonon branch is described in ref. \citep{Ruqing2005}; we follow a similar set of equations for our analysis.  

\begin{figure*}
    \centering
    \includegraphics{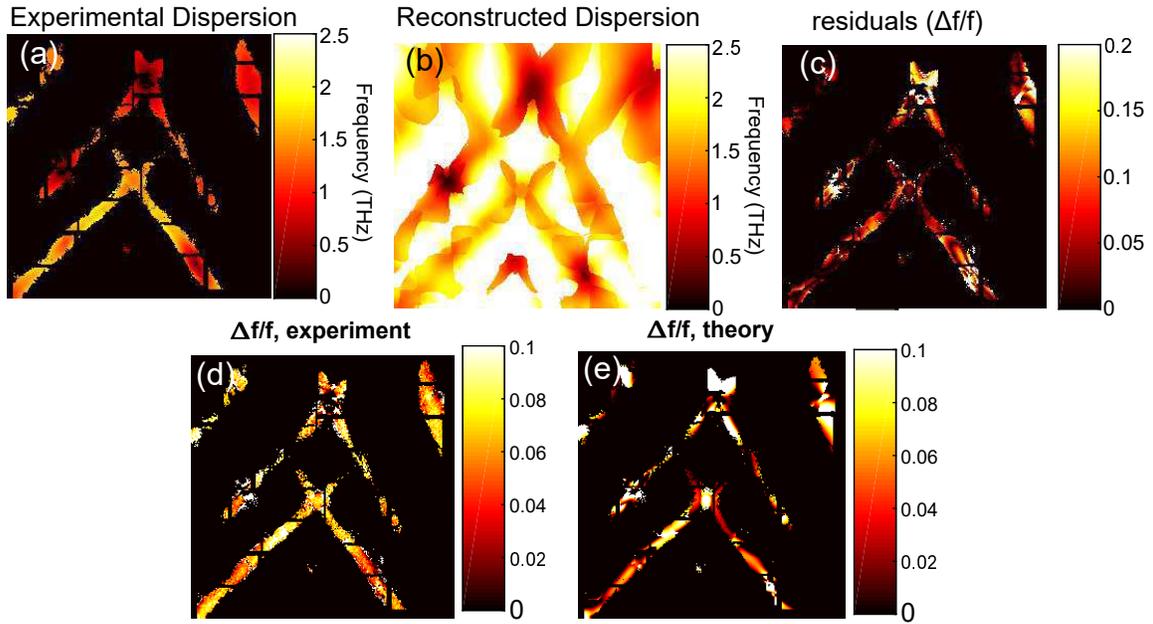}
    \caption{(a) Fitted frequencies, as in fig. \ref{fig:Bi_area_fitting_results}(a).  (b) Results of the frequencies from the force fitting. (c) Residuals (fractional) from 0 to 20\% of the frequencies relative to the fit.  (d) Same in fig. \ref{fig:Bi_area_fitting_results}(b).  (e) Theoretical reconstruction of (d).}
    \label{fig:Bi_area_fitting_forces}
\end{figure*}

The absolute value of the residuals for the fitting routine are shown in fig. \ref{fig:Bi_area_fitting_forces}(c).  The value varies from 0-20\% of the observed frequency, depending on wavevector.  A much higher precision fit can be obtained by only taking cuts along the high-symmetry directions, but at the cost of poor agreement away from high-symmetry directions.

Agreement between the reconstructed fit and experimentally observed frequency softening over reciprocal space is shown in fig. \ref{fig:Bi_area_fitting_forces}(d) and (e).  In these panels, the softening is shown to be relatively independent of wavevector, though slightly higher near the zone boundaries.  This is in rough agreement with  our fitted model.  

